\documentclass{aa}
\usepackage{epsf}

\begin{document}

  \thesaurus{12.          
              (08.14.1;   
               02.04.1)   
             }


\title{ Neutrino emission due to proton
        pairing in neutron stars}

\author{
        A.D.~Kaminker\inst{1}\and P.~Haensel\inst{2}\thanks{%
E-mail: haensel@camk.edu.pl}\and
        D.G.~Yakovlev\inst{1}
        }
\institute{
        Ioffe Physical Technical Institute, Politekhnicheskaya 26,
        194021 St.-Petersburg, Russia
\and
       N.~Copernicus Astronomical Center,
       Bartycka 18, 00-716 Warszawa, Poland
       }

\date{Received 3 March  1999 / Accepted  19 March 1999}

\offprints{P.~Haensel}

\titlerunning{Neutrino emission due to proton pairing}
\authorrunning{A.D.~Kaminker et al.}

\maketitle


\begin{abstract}
We calculate the neutrino
energy emission rate
due to singlet-state pairing of protons in the neutron star cores
taking into account the relativistic correction to
the non-relativistic rate.
The non-relativistic rate is numerically small, and  the
relativistic correction appears to be about 10 -- 50 times larger. 
It plays thus 
the leading role, reducing great difference
between the neutrino emissions due to
pairing of protons and neutrons. 
The results are important
for simulations of neutron star cooling.
\end{abstract}

Various mechanisms of neutrino emission in the cores
of neutron stars (NSs) are important
for NS cooling (e.g., Pethick \cite{p92}).
In this Letter, we analyze a
specific neutrino reaction
associated with superfluidity of protons.
The reaction consists in
neutrino-pair emission
$
          p \rightarrow p + \nu + \bar{\nu}
$
by protons whose dispersion relation contains superfluid
energy gap.
In theoretical studies,
this process is treated as annihilation of quasiprotons
$\tilde{p}$
into neutrino pairs,
$\tilde{p} + \tilde{p} \to \nu + \bar{\nu}$,
and is referred to as
{\it neutrino emission due to Cooper pair formation}.
Similar but more powerful process
takes place due to pairing of neutrons.

It is widely accepted (e.g., Takatsuka \& Tamagaki \cite{tt93}
and references therein), that nucleon superfluidity
in the NS cores occurs mainly due to triplet-state pairing of neutrons
and singlet-state pairing of protons.
The neutrino emission due to singlet-state pairing of neutrons
(as the simplest example)
has been calculated by Flowers et al.\ (\cite{frs76}) and
Voskresensky \& Senatorov (\cite{vs86}) while the cases of
triplet-state pairing of neutrons and singlet-state
pairing of protons have been studied by Yakovlev et al.\ (\cite{ykl99},
hereafter Paper I).
The importance of neutrino emission produced by nucleon pairing
has been demonstrated by simulations of NS cooling
(e.g., Schaab et al.\ \cite{svsww97}, Page \cite{p98}, Paper I).

So far the neutrino energy emission rate (emissivity)
due to pairing of nucleons
has been calculated in the approximation
of non-relativistic nucleons.
We will show that the emission due to singlet-state pairing of protons
is actually determined by the relativistic correction.

The process in question goes through neutral electroweak currents
and is accompanied by emission 
of neutrinos of all flavors.
As shown in Paper~I, the  non-relativistic
term in the neutrino emissivity
is produced by vector electroweak proton current
and can be written as ($\hbar = c = k_{\rm B} =1$)
\begin{eqnarray}
 Q_0 & = & {4 G_{\rm F}^2
           m_{\rm p}^\ast p_{\rm F} \over 15 \pi^5}
           \, T^7 \, {\cal N}_\nu a_0 \, F(y),
\label{Q0}
\end{eqnarray}
where $G_{\rm F}$ is the Fermi weak interaction constant,
$T$ is temperature,
$m_{\rm p}^\ast$ is an effective proton mass in dense matter
(determined by the proton density of states near the
Fermi level), $p_{\rm F}$
is the proton Fermi momentum, and
${\cal N}_\nu=3$ is the number of neutrino flavors. The function
\begin{equation}
    F(y)=y^2 \int_0^\infty {z^4 \, {\rm d}x \over ({\rm e}^z+1)^2},
\label{F}
\end{equation}
depends on
$y=\Delta/T$, where $\Delta$ is the proton superfluid gap;
$x=\eta/T$,
$z=E /T = \sqrt{x^2+y^2}$, 
$E$ is the quasiproton energy
with respect to the Fermi level,
$\eta=v_{\rm F}(p-p_{\rm F})$,
$v_{\rm F}=p_{\rm F}/m_{\rm p}^\ast$ is the proton Fermi velocity.
Finally,
$a_0=(1-4 \, \sin^2 \Theta_{\rm W})^2 \approx 0.0064$, and
$\Theta_{\rm W}$ is the Weinberg angle
($\sin^2 \Theta_{\rm W} = 0.23$). Numerical smallness of $a_0$
for protons
comes from their quark structure.
Accordingly, the neutrino emission due to
proton pairing is
greatly reduced as compared with the emission
due to neutron pairing; for instance,
one has $a_0=1$, for singlet-state pairing of neutrons.

Let us calculate the relativistic correction $Q_1$ 
to $Q_0$ determined by the axial-vector electroweak proton currents.
The corresponding interaction
Hamiltonian  is
%
   $  H_A = -  G_{\rm F}
     c_A \, J^\alpha l_\alpha /( 2 \sqrt{2}) ,$
%
where $J^\alpha = \overline{\psi}_{\rm p}\gamma^\alpha \gamma^5 \psi_{\rm p}$
is the axial-vector proton 4-current ($\alpha=$0, \ldots 3),
$l^\alpha = (\bar{\psi}'_\nu \gamma^\alpha (1 + \gamma^5) \psi_\nu)$
is the neutrino 4-current;
$\psi_{\rm p}$ is the bispinor wavefunction
of quasiprotons
(normalized with respect to one particle in unit volume),
$\psi_\nu$ and $\psi_\nu'$
are the standard bispinor amplitudes of emitted  neutrinos;
upper bar denotes a Dirac conjugate,
$\gamma^\alpha$ and $\gamma^5$ are Dirac matrices,
and $c_A$ is the axial-vector constant 
(for protons, $c_A= - g_A = -1.26$). 
The  bispinor $\psi_{\rm p}$ can be presented as
combination of an upper spinor $\phi_{\rm p}$ and lower spinor $\chi_{\rm p}$.
Writing the 4-vector $J^\alpha$ as ($J^0$, {\vec J}), we have
$   
    J^0= - \phi_{\rm p}^+ \chi_{\rm p} - \chi_{\rm p}^+ \phi_{\rm p},$ and
  $ {\vec J} = - \phi_{\rm p}^+ {\vec \sigma} \phi_{\rm p} - 
\chi_{\rm p}^+ {\vec \sigma} \chi_{\rm p},
$
where ${\vec \sigma}$ is the Pauli matrix.

Let $\hat{\Psi}$ be the non-relativistic 
second--quantized Bogoliubov spinor wave function
of quasiprotons in superfluid matter
(Lifshitz \& Pitaevskii 1980).
For our purpose, it is sufficient to set
$\phi_{\rm p}= \hat{\Psi}$,
$\chi_{\rm p}= -i {\vec \sigma} \nabla \hat{\Psi}/(2 m_{\rm p})$,
where $m_{\rm p}$ is 
the bare proton  mass.
Since the lower spinor $\chi_{\rm p}$ is
small as compared to $\phi_{\rm p}$ we can neglect
the term
$\chi_{\rm p}^+ {\vec \sigma} \chi_{\rm p}$ in {\vec J}.
Further considerations  are  similar to those  in Paper~I,
and we omit the details.

The squared matrix elements of the interaction
Hamiltonian summed over quasiproton spin states
contain the tensor components
$I^{\alpha \beta}  =  \sum_{\rm spins} \,
   \langle B | J^\alpha | A \rangle \,
   \langle B | J^\beta | A \rangle^\ast$,
where $| A \rangle$ and $| B \rangle$ denote,
respectively, initial and final states of the quasiproton system
(e.g., Paper I). Direct calculation shows that
the tensor $I^{\alpha \beta}$ is diagonal, 
with two nontrivial
matrix elements:
\begin{eqnarray}
   I_{\rm t} & \equiv & I^{00} =  (u'v+v'u)^2
                     |{\vec p}-{\vec p}'|^2/(2 m_{\rm p}^2),
\nonumber \\
   I_{\rm s} & \equiv & I^{11}  =  I^{22} = I^{33} = 2 (u'v-v'u)^2,
\label{II}
\end{eqnarray}
where $u= \sqrt{(E+\eta)/(2E)}$ and $v=\sqrt{(E-\eta)/(2E)}$
are the coefficients of the Bogoliubov transformation
for an annihilating quasiproton with momentum {\vec p}
and energy $E= \sqrt{\eta^2 + \Delta^2}$,
while $u'$ and $v'$ are the same coefficients for
a second annihilating quasiproton with momentum ${\vec p}'$
and energy $E'$.
The general expression for the neutrino emissivity $Q_1$
due to axial-vector proton interaction $H_A$
is readily given by Eq.\ (13) or (16) of Paper I, where
$c_V$ has to be formally replaced by $c_A$. Taking into
account diagonality of $I^{\alpha \beta}$ we obtain
\begin{eqnarray}
   Q_1 & = & \left( {G_{\rm F} \over 2 \sqrt 2} \right)^2 \,
       {2 \pi \over 3} \, { {\cal N}_\nu \over (2 \pi)^8} \, c_A^2 \,
       \int {\rm d} {\vec p} \;
       {\rm d} {\vec p}' \;
       f(E)f(E') \, \omega
\nonumber \\
  &  & \times  \left[  {\vec k}^2 I_{\rm t}
      + (3\omega^2 - 2{\vec k}^2) \, I_{\rm s} \right].
\label{Qgen1}
\end{eqnarray}
In this case, $\omega$ and ${\vec k}$ are, respectively, energy and momentum
of a neutrino pair
($\omega=E + E', {\vec k}={\vec p} + {\vec p}'$),
$f(E)=( {\rm e}^{E/T}+1)^{-1}$
is the Fermi-Dirac distribution;
integration over
{\vec p} and ${\vec p}'$
is done over
the kinematically allowed domain $\omega^2 > {\vec k}^2$.

Since the proton Fermi liquid in NS cores is strongly
degenerate,
we set $p=p'=p_{\rm F}$
in all smooth functions under the integral.
The presence of energy gaps
opens the reaction kinematically
in a small region of momentum space where
{\vec p} is almost antiparallel to ${\vec p}'$. This
enables us to set ${\vec p}'=-{\vec p}$
in all smooth
functions in the integrand.
Then Eq.\ (\ref{II})
gives $I_{\rm t}= 2 (p_{\rm F}/m_{\rm p})^2 (\Delta/E)^2$
while $I_{\rm s} \to 0$
for  $p \to p'$.
Expanding $I_{\rm s}$ in the vicinity of $p=p'$ we obtain
$I_{\rm s}= v_{\rm F}^2 \Delta^2\,(p-p')^2/(2 E^4)$.
After some transformations
described in Paper I
we come to Eq.\ (20) of Paper I in which again
$c_V$ should be replaced by $c_A$.
Subsequent integration over $p$ and $p'$
is quite analogous, and finally we come to
the expression for $Q_1$ which can be obtained from (\ref{Q0}) by  
replacing $a_0 \to a_1$, where
\begin{eqnarray}
          a_1 & = & c_A^2 v_{\rm F}^2 \, \left[
          ( m_{\rm p}^\ast / m_{\rm p} )^2 + 11/ 42 \right].
\label{a1}
\end{eqnarray}
As seen from the derivation, both non-trivial tensor
components, $I_{\rm t}$ and $I_{\rm s}$,
contribute to $Q_1$, producing the first and second
terms in square brackets in Eq.\ (\ref{a1}), respectively.
The relativistic correction to the neutrino
emissivity due to singlet-state pairing of neutrons
has been calculated by Flowers et al.\ (\cite{frs76}).
It should be the same for neutrons and
protons since the squared axial-vector constant $c_A^2=g_A^2$
is the same. However, the
expression obtained by Flowers et al.\ (\cite{frs76})
contains only the second term in square brackets in (\ref{a1}).
Clearly the first term was overlooked.

Introducing $Q=Q_0+Q_1$ and $a=a_0+a_1$ we
obtain the total neutrino emissivity
(in the standard physical units)
\begin{eqnarray}
 Q& =  &   {4 G_{\rm F}^2
           m_{\rm p}^\ast p_{\rm F} \over 15 \pi^5 \hbar^{10}
           c^6} \, (k_{\rm B} T)^7 \, {\cal N}_\nu \, a \, F(y)
  =  1.170 \times 10^{21}
 \nonumber \\
   &  & \times
    \left( {m_{\rm p}^\ast \over m_{\rm p} } \right)^2 
       { v_{\rm F} \over c   } \,
       \, T_9^7 \, {\cal N}_\nu \, a \, F(y)~~~
      {{\rm erg} \over {\rm cm^3 \; s  }},
\label{QQ} \\
a & = &    0.0064 +
    1.60 \, \left(  v_{\rm F}/ c  \right)^2
       \left[ \left( m_{\rm p}^\ast / m_{\rm p} \right)^2 + 11/ 42 \right].
\label{a}
\end{eqnarray}
Let us remind a useful  relationship
$v_{\rm F}/c=$ 0.353 $(m_{\rm p}/m_{\rm p}^\ast) \,(n_{\rm p}/n_0)^{1/3} $, 
where $n_{\rm p}$ is the proton number density and
$n_0=0.16$ fm$^{-3}$ is saturation nuclear matter density.

One can see that
the relativistic correction greatly exceeds
the non-relativistic term under typical conditions prevailing 
in the NS cores.
Notice also that Eqs.\ (\ref{QQ}) and
(\ref{a}) neglect the relativistic corrections to $Q$
due to vector proton currents.
\footnote{
We have shown that inclusion of these corrections leads
formally to Eq.\ (1), where $F$ is given by Eq.\ (2)
but with additional factor $C$ under integral: $C$=$1+v_{\rm F}^2 \, D/z^2$, 
$D$=$-5z y^2/21$ $ +(10z/63)(y^2-3x^2)/({\rm e}^z+1)+ 
(10/21)(z^2+y^2)-(zm_{\rm p}^*/m_{\rm p})^2$. 
}
This is justified since
the main term, 
in  our case, is numerically
small and unimportant.
In the cases of any neutron pairing or triplet-state
proton pairing the  
non-relativistic term
is not numerically small and
all relativistic corrections can be neglected, 
in practice.

The fit expression of $F(y)$ is (Paper I)
\begin{eqnarray}
  F(y) & = &  (0.602 \, y^2 + 0.5942\, y^4 +
     0.288 \, y^6)
 \nonumber \\
       & & \times 
     \left( 0.5547 + \sqrt{(0.4453)^2 + 0.01130 \,y^2} \right)^{1/2}
\nonumber \\
       &  & \times 
     \exp \left(- \sqrt{4 \, y^2 + (2.245)^2 } + 2.245 \right);
\label{fit}\\
  y &  = &
   \sqrt{1 - \tau }
        \left(1.456 -  0.157 \, \tau^{-1/2}  +
         1.764 \, \tau^{-1} 
         \right),
\nonumber
\end{eqnarray}
where
$\tau =T/T_{\rm cp}<1$, and
$T_{\rm cp}$ is the
superfluid critical temperature
of protons.
Equations (\ref{QQ}) --
(\ref{fit}) enable one
to calculate the neutrino emissivity $Q$.

%
\begin{figure}[t]
\begin{center}
\leavevmode
\epsfysize=7.5cm
\epsfbox{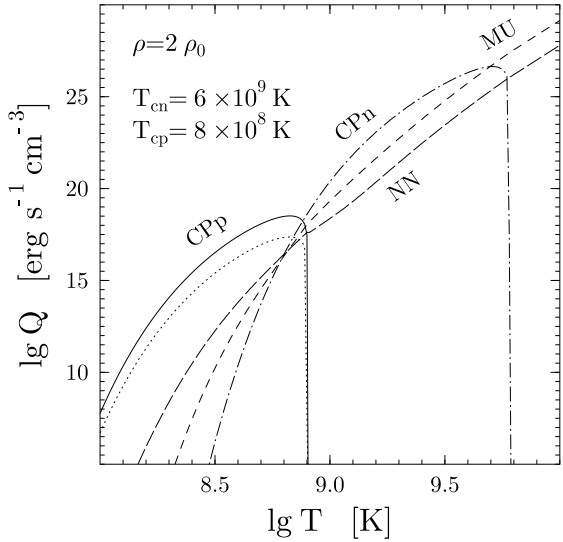}
\end{center}
\vspace{-0.7cm}
\caption[]{
     Temperature dependence of
     neutrino emissivities
     in $npe$ matter at $\rho = 2 \, \rho_0$.
     The reactions are: modified Urca
     (MU, sum of $n$ and $p$ branches),
     nucleon-nucleon bremsstrahlung (NN, sum of
     $nn$, $np$ and $pp$ branches),
     Cooper pairing of neutrons (CPn) and protons (CPp). The
     emissivity  from the 
     latter reaction is shown twice, in the non-relativistic
     approximation (dots), and with the relativistic
     correction included (solid line).
     }
\label{fig1}
\end{figure}

Let us discuss the importance of neutrino emission
due to Cooper pairing of protons
in superfluid NS cores. For illustration, we will adopt
a moderately stiff equation of state
proposed by Prakash et al.\ (\cite{pal88})
(the same version as used in Paper I).
We assume that dense matter consists of
neutrons, protons and electrons (no muons and hyperons).
The equation of state allows the direct Urca process
to operate at $\rho \geq 4.64 \, \rho_0$, 
where $\rho_0= 2.8 \times 10^{14}$ g cm$^{-3}$ is the saturation
nuclear matter density.
Consider two densities, $\rho=2 \, \rho_0$ and
$\rho= 5 \, \rho_0$, as examples of the standard neutrino
cooling (direct Urca is forbidden)
and the cooling enhanced by the direct Urca process, respectively.
The nucleon effective masses will be set equal to $0.7$
of the masses of bare particles.
Neutrons and protons
are supposed to form superfluids
(triplet-state neutron
and singlet-state proton pairing)
with critical temperatures $T_{\rm cn}$ and $T_{\rm cp}$, respectively.

We compare the 
neutrino emissivity due to pairing of protons with 
those resulting from the main neutrino reactions in superfluid NS
cores: direct and modified Urca reactions, nucleon-nucleon
bremsstrahlung (as described in  Yakovlev \& Levenfish \cite{yl95}
and Levenfish \& Yakovlev \cite{ly96}),
and Cooper pairing of neutrons
(Paper I).
Our analysis shows that the emission due to pairing of protons
is especially important for
$T \la T_{\rm cp} \ll T_{\rm cn}$. 
In this case, strong neutron
superfluidity 
greatly suppresses
almost all other neutrino
reactions
and Cooper pairing of protons becomes dominant.

Figure 1
shows temperature dependence of the neutrino emissivities
at $\rho = 2 \rho_0$,  $T_{\rm cn}=6 \times 10^9$ K and
$T_{\rm cp}= 8 \times 10^8$ K.
For  $T > T_{\rm cn}$ matter is nonsuperfluid,
and the modified Urca reaction dominates.
For lower $T$, neutrino emission due to neutron pairing becomes most
important but it dies out at $T \la 0.1 \, T_{\rm cn}$.
The emission due to proton pairing 
dominates at $T \la T_{\rm cp}$. 
Including
the relativistic correction increases its efficiency
by a factor of about 10. 

%
\begin{figure}[t]
\begin{center}
\leavevmode
\epsfysize=7.5cm
\epsfbox{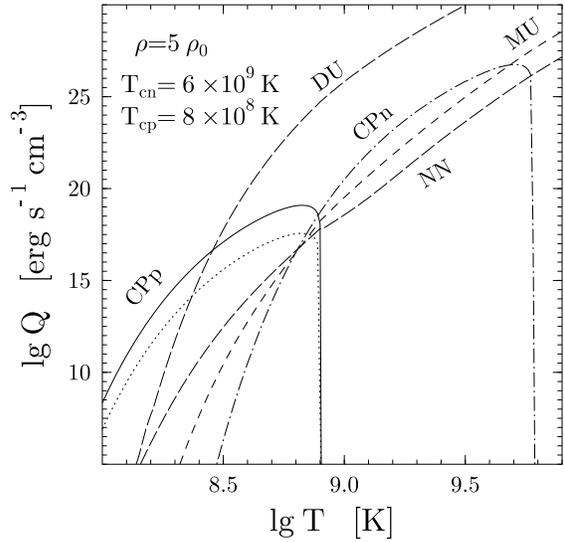}
\end{center}
\vspace{-0.7cm}
\caption[]{
     Same as in Fig.\ 1 but for
     $\rho = 5 \, \rho_0$.
     Additionally,  direct Urca process (DU) is allowed.
     }
\label{fig5}
\end{figure}

Figure 2 is the same as Fig.\ 1
but for $\rho=5 \, \rho_0$. Now
 the powerful direct Urca process is allowed and
dominates at
$T \ga 3 \times 10^8$ K. However, in spite of its high
efficiency in nonsuperfluid matter, it is strongly suppressed
at lower $T$, so that the neutrino emission due to
proton pairing becomes dominant.
The relativistic correction
enhances this process stronger
than at $\rho = 2 \, \rho_0$,
by a factor $\sim$ 30.

Concluding, 
the relativistic correction greatly enhances 
the neutrino emission due to proton pairing
and brings it
closer to the neutrino production
due to neutron pairing. This
is important for
NS cooling. 

\begin{acknowledgements}
Two of the authors (ADK and DGY) acknowledge
hospitality of N.\ Copernicus Astronomical
Center. 
This work was supported in part by the
RBRF (grant No. 99-02-18099), INTAS (grant No. 96-0542),
and KBN
(grant 2 P03D 014 13).
\end{acknowledgements}

\end{document}